# Performance of STBC MC-CDMA systems over outdoor realistic MIMO channels

F. Portier, J-Y. Baudais, J-F. Hélard
IETR, UMR CNRS 6164
INSA, 20 avenue des Buttes-de-Coësmes, CS 14215, 35043 Rennes Cedex, FRANCE
{fabrice.portier, jean-yves.baudais, jean-francois.helard}@insa-rennes.fr

*Abstract*—This paper deals with orthogonal Space-Time Block Coded MC-CDMA systems, in outdoor realistic downlink scenarios with up to two transmit and receive antennas. Assuming no channel state information at transmitter, we compare several linear single-user detection and spreading schemes, with or without channel coding, achieving a spectral efficiency from 1 to 2 bits/s/Hz. The different results obtained demonstrate that spatial diversity significantly improves the performance of MC-CDMA systems, and allows different chip-mapping without notably decreasing performance. Moreover, the global system exhibits a good trade-off between complexity at mobile stations and performance. Then, Alamouti's STBC MC-CDMA schemes derive full benefit from the frequency and spatial diversities and can be considered as a very realistic and promising candidate for the air interface downlink of the 4$^{th}$ generation mobile radio systems.

## I. INTRODUCTION

The downlink of future cellular mobile radio systems will call for high throughput services. Such a substantial multi-user datarate on a limited bandwidth means high spectral efficiency from the physical layer. To benefit from the channel characteristics like fading diversity and to approach maximal capacity with a reasonable complexity at receiver in the downlink, several techniques are studied for Beyond-3G systems. When no Channel State Information (CSI) is available at transmitter, spreading information in time, frequency and space to mitigate fading effect is advisable. Promising techniques are then efficient Channel Coding (CC), Multi-Carrier Code Division Multiple Access (MC-CDMA) and Multiple-Input Multiple-Output (MIMO) systems.

MC-CDMA modulation scheme has already proven to be a strong candidate as an access technique for the downlink of broadband cellular systems [1] [2] [3], as it benefits from the advantages of both MC and CDMA techniques: high spectral efficiency, robustness in frequency selective channels with a low-complexity at receiver considering a simple one-tap equalization, multiple access capability with high flexibility, narrow-band interference rejection... Thus, MC-CDMA is for example studied within the European IST-MATRICE project. This work has been partly carried out within this project which aims at defining a new air interface for 4G systems.

However, such techniques combined with efficient channel turbo-coding can nearly reach the limited Single-Input Single-Output (SISO) channel capacity [9]. Thus, to increase this capacity limit and to benefit from spatial diversity, we study a synchronous downlink system with several antennas at transmitter (Base Station BS) and also at receivers (Mobile Stations MS). We first consider MISO (Multiple-Input Single-Output) and then MIMO (Multiple-Input Multiple-Output) transmissions, with Space-Time Block Coding (STBC) as we assume no CSI at transmitter.

Even if capacity is the main advantage of MIMO systems when increasing the number of antennas at both sides (BS and MS), we focus on a unitary-rate STBC as we target a mobile receiver with two but also only one antenna, and we want a system operational in spatially correlated channels. In particular, we consider MC-CDMA with Alamouti's STBC [5] [6] systems using Zero Forcing (ZF) and Minimum Mean Square Error (MMSE) Single-user Detection (SD) schemes. Given the channel characteristics, we apply Alamouti coding in space and time, per subcarrier. In that case, Alamouti scheme provides full spatial diversity gain, no intersymbol interference and low complexity Maximum Likelihood receiver as transmission matrix is orthogonal. Moreover with that orthogonal STBC, only one receive antenna can be used. In such a MISO design, with two transmit antennas and one receive antenna, Alamouti's STBC is also optimal for capacity. Furthermore, we will not study STBC with rate inferior to one, because the reduction in coding rate is best invested in turbo-codes than in STBC, when combined with efficient channel turbo-coding.

In this paper, we compare the performance of Alamouti's STBC MC-CDMA systems for outdoor scenarios, using a realistic MIMO channel model and configuration. This 3GPP2-like MIMO channel developed within the IST-MATRICE project is configured to fit a 57.6 MHz sampling frequency for realistic outdoor propagations. We give inputs to see the use of diversity in space, time and frequency, stressing the compromise with users' orthogonality and showing its effect on Bit Error Rates (BER) or Frame Error Rates (FER) results.

This paper is organized as follows. Section 2 describes the studied orthogonal STBC MC-CDMA system. Section 3 deals with system parameters, including MIMO channel configuration for outdoor, the system choices and inputs on diversity exploitation. Section 4 gives the performance of STBC MC-CDMA systems with ZF and MMSE SD detection in that outdoor context, achieving a spectral efficiency from 1 to 2 bits/s/Hz.

## II. SYSTEM DESCRIPTION

### A. General presentation

Figure 1 shows a simplified MIMO MC-CDMA system for user $j$ based on Alamouti's STBC with $N_t = 2$ transmit antennas and $N_r = 2$ receive antennas. Channel coding or framing processes are omitted. Each user $j$ simultaneously transmits symbols $x_j^1(u) = s_j^1$ and $x_j^2(u) = s_j^2$ from both antennas at time $u$, then $-s_j^{2*}$ and $s_j^{1*}$ at time $u+T_x$ where $T_x$ is the OFDM symbol duration. Dropping time index at the space-time encoder output, the data symbols of the $N_u$ users $\mathbf{x}^1 = [x_1^1 \ldots x_j^1 \ldots x_{N_u}^1]^T$ are multiplied by their specific orthogonal spreading code $c_j = [c_{j,1} \ldots c_{j,Lc}]^T$ where $c_{j,k}$ is the $k^{th}$ chip, and $[.]^T$ denotes matrix transposition (the same goes for symbol $\mathbf{x}^2$). $c_j$ is the $j^{th}$ column vector of the $L_c \times N_u$ spreading code real-valued matrix $C$. Fast Hadamard Transform can be used in downlink to spread and sum data of all users. Note that we can swap the linear processes (STBC and spreading). For outdoor scenarios, the length $L_c$ of spreading sequences is inferior to the number $N_c$ of subcarriers. Thus, depending on the diversity and orthogonality degrees we want to favor, different chip mapping (linear frequency interleaving or 2D-spreading for example) can be carried out. This compromise is explained in part III. In the following equations, we consider only $L_c$ subcarriers and 1D-spreading in frequency (classical MC-CDMA) without losing generality as the extension (index rearrangement) is straightforward. Each data symbol is then transmitted on $L_c$ parallel subcarriers. The vector obtained at the $r^{th}$ receive antenna after the OFDM demodulation and deinterleaving, at time $u$ and $u+T_x$, is given by

$$Y_r = H_r DS + N_r \quad \text{with} \quad H_r = \begin{bmatrix} H_{1r} & H_{2r} \\ H_{2r}^* & -H_{1r}^* \end{bmatrix} \quad (1)$$

where $Y_r = [y_r^T(u) \, y_r^H(u+T_x)]^T$ with $y_r(u) = [y_{r,1}(u) \ldots y_{r,Lc}(u)]^T$ the vector of the $L_c$ received signals at time $u$ and $[.]^H$ denotes the Hermitian (or complex conjugate transpose); $H_{tr} = \text{diag}\{h_{tr,1}, \ldots, h_{tr,Lc}\}$ ($t,r \in \{1,2\}$) is a $L_c \times L_c$ diagonal matrix with $h_{tr,k}$ the complex channel frequency response, for the subcarrier $k$ from the transmit antenna $t$ to the receive antenna $r$. Time invariance during two MC-CDMA symbols is assumed to permit the recombination of symbols when STBC is used in time. $D = \text{diag}\{C,C\}$ and $S = [s^{1T} \, s^{2T}]^T$. $N_r = [n_r^T(u) \, n_r^H(u+T_x)]^T$ where $n_r(u) = [n_{r,1}(u) \ldots n_{r,Lc}(u)]^T$ is the Additive White Gaussian Noise (AWGN) vector with $n_{r,k}(u)$ representing the noise term at subcarrier $k$, for the $r^{th}$ receive antenna at time $u$ with variance given by $\sigma_k^2 = E\{|n_{r,k}|^2\} = N_0 \, \forall k$.

### B. Linear Single-user Detection in MIMO

At the receiver side, in order to detect the two transmitted symbols $s_j^1$ and $s_j^2$ for the desired user $j$, ZF or MMSE SD schemes are applied to the received signals in conjunction with STBC decoding. In the SISO case, MMSE SD is the most efficient linear SD scheme [4], and thus given as reference to evaluate the gain of MISO and MIMO systems. In the MIMO case, after equalization for each receive antenna, the two successive received signals are combined and added from the $N_r$ receive antennas to detect the two symbols.

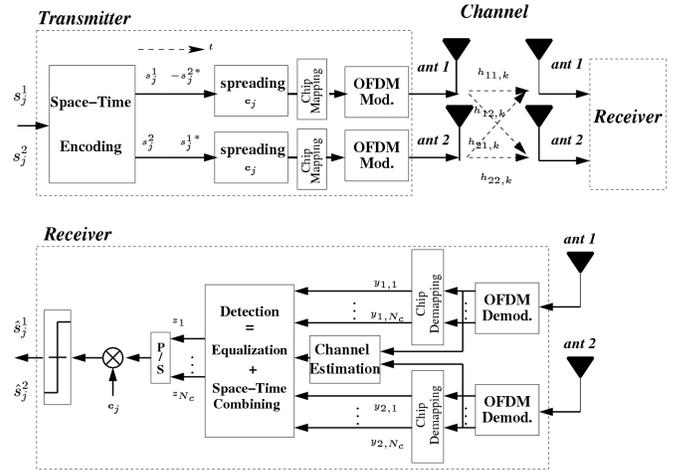

Figure 1. MC-CDMA transmitter and receiver for user $j$ with transmit and receive diversities.

After despreading and threshold detection, the data symbols $\hat{s}_j^1$ and $\hat{s}_j^2$ for user $j$ are

$$\left[\hat{s}_j^1 \, \hat{s}_j^2\right]^T = (\mathbf{I}_2 \otimes c_j^T) \mathbf{Z} = (\mathbf{I}_2 \otimes c_j^T) \sum_{r=1}^{N_r} \mathbf{G}_r \mathbf{Y}_r \text{ with } \mathbf{G}_r = \begin{bmatrix} \mathbf{G}_{1r} & \mathbf{G}_{2r}^* \\ \mathbf{G}_{2r} & -\mathbf{G}_{1r}^* \end{bmatrix} \quad (2)$$

where $\mathbf{I}_2$ is the 2x2 identity matrix, $\otimes$ the Kronecker product, $\mathbf{Z} = [z_1^1 \ldots z_{Lc}^1 \, z_1^2 \ldots z_{Lc}^2]^T$ the vector of the received signals equalized and combined from the $N_r$ antennas, and $\mathbf{G}_{tr}$ a diagonal matrix (since we used an SD scheme) containing the equalization coefficients for the channel between the transmit antenna $t$ and the receive antenna $r$. For instance, to detect $s_i^1$, the MMSE SD coefficients $g_{tr,k}$ minimize the mean square value of error $\varepsilon_k^1$ between the signal $\sum_{i=1}^{N_u} c_{i,k} s_i^1$ transmitted on subcarrier $k$ and the received signals combined from the $N_r$ receive antennas by the Alamouti decoding. In the same way, the ZF coefficients $g_{tr,k}$ restore the orthogonality between the different users. It is well known that with SISO systems, ZF leads to excessive noise amplification for low subcarrier Signal to Noise Ratio (SNR). In the MIMO case, spatial diversity, equal to the product $N_t \times N_r$ in the decorrelated situation, statistically reduces this occurrence. Thus, with an increasing number of antennas, ZF tends to MMSE efficiency, and does not require a SNR estimation of $\gamma$ at receiver. We assume the same noise level statistically for each subcarrier or receive antenna. Besides, knowledge of the spreading codes $c_i (i \neq j)$ of the interfering users is not required to derive the ZF and MMSE Single-user Detection coefficients, as shown in the following MMSE equation:

$$g_{tr,k} = h_{tr,k}^* / \left[ \sum_{t=1}^{2} \sum_{r=1}^{N_r} |h_{tr,k}|^2 + \frac{1}{\gamma} \right] \quad (3)$$

ZF equations are similar assuming $1/\gamma = 0$. Note that the threshold detection should be normalized by $\rho$ for MMSE with high-order modulations (16QAM...). The sum is performed on the $L_c$ subcarriers where is spread the considered symbol:

$$\rho = L_c / \sum_{k=1}^{L_c} \frac{\sum_{t=1}^{2}\sum_{r=1}^{N_r} |h_{tr,k}|^2}{\sum_{t=1}^{2}\sum_{r=1}^{N_r} |h_{tr,k}|^2 + \frac{1}{\gamma}} \quad (4)$$

III. SYSTEM PARAMETERS

*A. MIMO channel configuration*

We use a link level MIMO channel model which has been specifically developed within the European research IST MATRICE project. This channel model is based on the 3GPP/3GPP2 proposal [8] for a wideband MIMO channel exploiting multipath angular characteristics. It consists in elaborating a spatial model from a hybrid approach between a geometrical concept depending on cluster positions and a tapped delay line model describing the Average Power Delay Profile (APDP) with a fixed number of taps. The spatial parameters are defined at the Base Station BS and at the Mobile Stations MS. The angular distribution is modeled from parameters leading to an average angle spread. The model parameters have been adapted at 5 GHz for outdoor environment. Then, the MIMO channel response consists in a sum of sub-rays extracted from the previous statistics.

Table I summarizes the main MIMO channel parameters of our outdoor propagation scenario. Outdoor is characterized by a large delay spread, high mobility, and an asymmetrical antenna configuration. The consequent spatial correlation is inferior to 0.1 for an antenna spacing of 10 λ (λ is wavelength), while a 0.5 λ spacing leads to a correlation around 0.7 at BS, and 0.35 at MS. These spatial characteristics are shown on Figure 2, where the circles reveal the angular distributions of sub-rays for a given channel realization, and the curves expose the resultant spatial correlation in function of antenna spacing at BS and MS respectively, averaged on 32 channel realizations. We use the BRAN E channel APDP [7] to model an outdoor urban environment. A measure of the coherence bandwidth (mono-sided at 3dB) is around 1.5 MHz, close to the theoretical BRAN E one. Correlation in time, derived from Doppler frequency (278 Hz for the considered velocity), is given by the measured coherence time that is close to the frame duration. Unless stated, we will present results with an antenna spacing of 10 λ at BS and 0.5 λ at BS.

TABLE I. MAIN MIMO CHANNEL PARAMETERS

| Channel Profile | BRAN E |
|---|---|
| Maximum delay $\tau_{max}$ | 1.76 µs |
| Number of paths | 17 |
| Number of sub-rays par path | 20 |
| Velocity | 60 km/h |
| Mean Angle Spread at BS | $E(\sigma_{AS}) = 21.4°$ |
| Mean Angle Spread at MS | $E(\sigma_{AS, MS}) = 68°$ |
| Mean total RMS Delay Spread | $E(\sigma_{DS}) = 0.25$ µs |
| Element spacing | 0.5 λ - 10 λ |

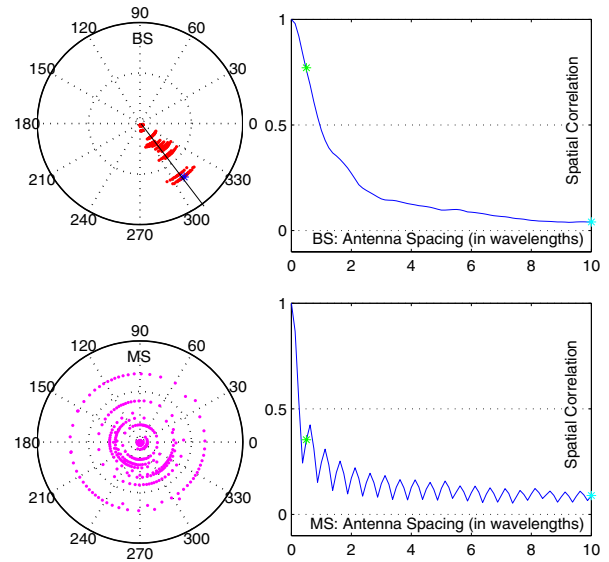

Figure 2. Spatial characteristics of our outdoor channel configuration.

*B. Main system configuration*

The system parameters are chosen according to the time and frequency coherence of the channel in order to reduce inter-subcarrier interference and Inter-Symbol Interference (ISI). Besides, investigated MC-CDMA configurations are designed to propose high throughput solutions for outdoor scenarios, as shown in Table II.

TABLE II. MAIN SYSTEM PARAMETERS.

| Sampling frequency Fs | 57.6 MHz |
|---|---|
| FFT size | 1024 |
| Number $N_c$ of used sub-carriers | 736 |
| Guard Interval duration $T_g$ | 3.75 µs |
| Total symbol duration $T_x = T_u + T_g$ | 21.52 µs |
| Subcarrier spacing $\Delta_f = 1/T_u$ | 56.2 KHz |
| Length $L_c$ of spreading codes | 16 - 32 |
| Modulation | QPSK – 16QAM |
| Center frequency $f_c$ | 5.0 GHz |
| Occupied bandwidth $B_f$ | 41.46 MHz |
| Frame duration / Guard duration | 30 $T_s$ / 20.8 µs |

The studied configuration proposed for outdoor scenarios is based on a sampling frequency which is a multiple of the 3.84 MHz UMTS frequency, to obtain the same frame duration as UMTS (0.666 ms). So $F_s$ is equal to 15 x 3.84 = 57.6 MHz. We consider a carrier frequency $f_c$ = 5 GHz, an FFT size of 1024 with $N_c$ = 736 used sub-carriers. The guard interval duration $T_g$ = 3.75 µs, chosen according to the maximum delay $\tau_{max}$ = 3.5 µs to avoid ISI, leads to a 18 % spectral efficiency loss and a power efficiency loss equal to 0.84 dB. The overall bit-rate is then 67 Mbits/s for QPSK without channel coding, shared between users. Furthermore, the length $L_c$ of the spreading codes is equal to 32, and a chip mapping is applied to the symbols before OFDM modulation.

Results are presented assuming perfect channel estimation at receiver. In the $E_b/N_0$ value, we take into account the system parameters (like the coding rate), except the guard interval efficiency loss and prospective framing overhead. We assume a unitary total transmit power, so MIMO results include the 3dB gain when duplicating the receive antennas.

*C. Diversity versus orthogonality*

From system parameters and coherence time, we can define a temporal diversity order that is inferior to 2 over a frame when velocity is inferior to 80 km/h. In the same way, we can verify a frequency diversity order proportional to the ratio of full bandwidth over coherence bandwidth, which is approximately 15 times the diversity in time.

We define a chip as a specified location in the frame among the $N_c$ subcarriers in frequency and the $N_t$ OFDM symbols in time. Each symbol information is spread over $L_c$ chips. The usual MC-CDMA approach considers 1D-spreading in frequency. As $L_c < N_c$, a linear frequency interleaving can be applied on each spread symbol so as to benefit from the frequency diversity over the whole bandwidth. On the other hand, this diversity decreases the orthogonality between codes and increases Multiple Access Interference (MAI). Thus we will consider two schemes whether no interleaving or frequency interleaving is performed, respectively named 1Da and 1Db. Going further, we can apply a 2D-spreading matching the channel characteristics in time and frequency to increase orthogonality (2Da, on a square block of adjacent subcarriers and OFDM symbols) or diversity (2Db). We define $L_c = S_f \times S_t$, where $S_f$ and $S_t$ are respectively the spreading lengths in frequency and time, with $S_t \neq 1$ when 2D-spreading is applied. Adding the spatial dimension, not only the CDMA requires orthogonality for correct decoding, but also the Alamouti Space-Time decoding. We first have to favor a constant channel for the two chips used in the Alamouti algorithm, and then we have to find a trade-off between orthogonality and diversity for CDMA spreading of a symbol, to finally exploit the remaining diversity at the bit level between symbols using interleaving and channel coding over the frame. In our outdoor scenarios, we apply Alamouti along the time axis, which is more correlated than frequency dimension. Then, 1D or 2D-spreading is performed according to the expected correlations, explaining the 2Da scheme on figure 3. For example, when channel variation in time is lower than in frequency, a snake in time seems the best solution to order the chips of a spread symbol favoring orthogonality.

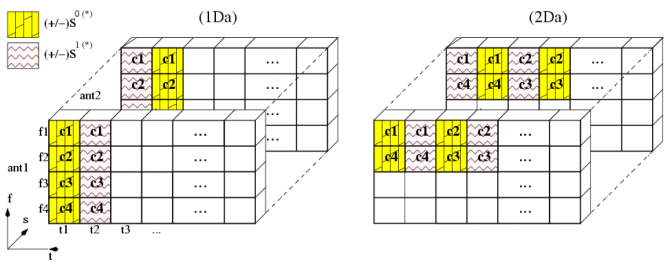

Figure 3. Chip-mapping examples, over time/frequency/space, assuming $S_f=N_c=4$, and 2 transmit antennas.

## IV. SIMULATION RESULTS

We present results in full-load, as it is the "worst-case" and we assume that the network can contrive (assigning several codes per user...) to fully exploit the provided throughput. Moreover with ZF detection, results are the same whatever the load. As the results and conclusions are similar in 1D-spreading and 2D-spreading, we mainly consider 1D-spreading ones.

*A. Performance using a QPSK input constellation*

Figure 4 shows the performance of the system in a pessimistic case, *i.e.* considering high spatial correlation using an antenna spacing of 0.5 λ only, at both sides. We employ QPSK symbols $s_j$ without channel coding, resulting in a theoretical spectral efficiency of 2 bits/s/Hz. Even with quite important spatial correlation, the spatial dimension gain (for MISO 21 or MIMO 22 schemes) is clear. Figure 5 considers a realistic case for correlations. We assume an antenna spacing of 10 λ at BS and 0.5 λ at MS. Spatial diversity still increases performance, especially for schemes that exploited less diversity in the previous figure like MISO and ZF. We mainly employed linear frequency interleaving (1Db).

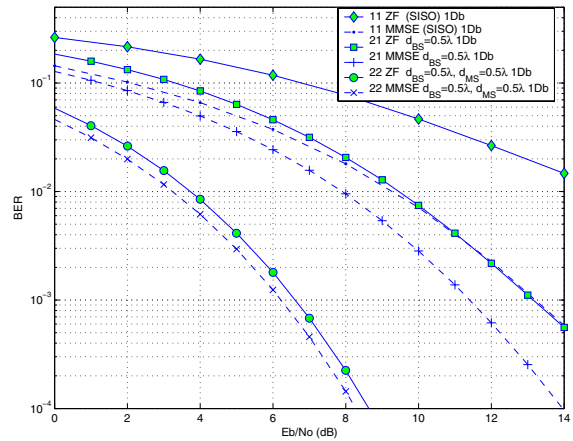

Figure 4. performance with QPSK and no channel coding at full-load ($N_u=L_c=32$) over a stronly spatially correlated outdoor channel.

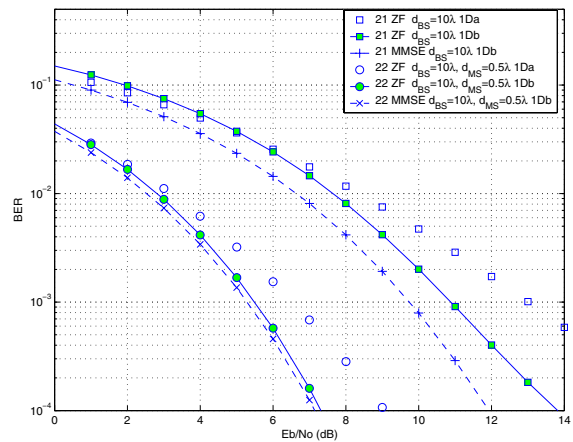

Figure 5. performance with QPSK and no channel coding at full-load ($N_u=L_c=32$) over a realistic outdoor channel.

Without channel coding to recover a part of the frame diversity for each transmitted bit, the chip mapping that favor orthogonality instead of diversity are disadvantaged. Best results are obtained with frequency interleaving (1Db or 2Db).

Figure 6 presents results obtained in a realistic configuration with channel coding and a 1 bit/s/Hz spectral efficiency (without considering guard-interval loss) *i.e.* around 33 Mbits/s for 32 users. The channel coding scheme is the turbo-code defined for UMTS, with a rate of 1/3, combined with a puncturing process and an interleaver to have an overall coding rate of 1/2. Results are given for 6 iterations at the decoder, and show that the global system benefits from spatial diversity. We present results without (1Da) or with (1Db) frequency interleaving. With channel coding, results are quite similar whatever the chip mapping, and favoring orthogonality still improves ZF detection (since it does not treat diversity like MMSE), tending to MMSE performance. In MISO case, it confirms that a simple STBC with unitary-rate equal to the channel rank, while coding is performed at bit level with the channel coder, is a good trade-off between complexity and performance for realistic scenarios.

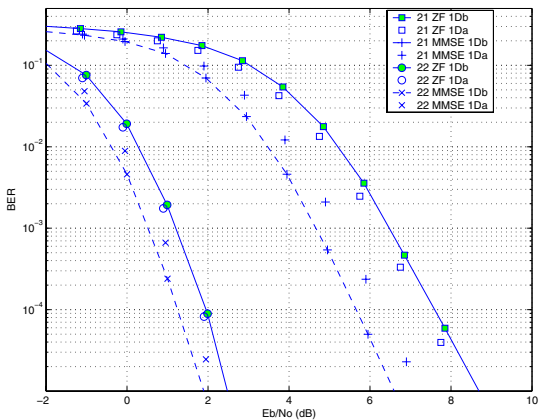

Figure 6.  performance with QPSK and UMTS-like turbo-coding at full-load ($N_u=L_c=32$) over a realistic outdoor channel.

*B. Performance in a higher datarate context*

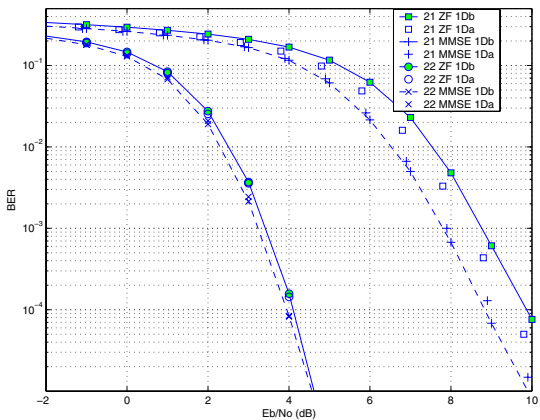

Figure 7.  Performance with 16QAM and UMTS-like turbo-coding at full-load ($N_u=L_c=32$) over a realistic outdoor channel.

Figure 7 shows performance using the same configuration as on figure 6, except 16QAM input symbols, to aim a 2 bits/s/Hz spectral efficiency. We obtain a datarate of 67 Mbits/s/Hz for 32 users, achieving a BER<$10^{-5}$ (and FER<$10^{-3}$) with $E_b/N_0$ inferior to 5 dB and 11 dB for studied MISO and MIMO systems respectively. When considering a practical system, these configurations with a simple ZF or sub-optimal MMSE detection (*i.e.* with a fixed *γ*, or ZF with a proper fixed-point treatment to avoid division by near-zero values) promise excellent performance without requiring SNR estimation for equalization at receiver.

## V. CONCLUSION

The performance of Alamouti's STBC MC-CDMA systems has been compared with ZF and MMSE single-user detections over a realistic MIMO channel model, in the case of two transmit antennas and one or two receive antennas. It demonstrates that spatial diversity significantly improves the performance of MC-CDMA systems, including in outdoor scenarios, and presents a good trade-off between performance and complexity. We pointed out the enhanced flexibility in chip-mapping, without conveying significant performance degradation. Thus, one of these schemes may be more efficient when considering impairments like real channel estimation. It allows us to have a simple reference system, with a rather low complexity, for further studies to propose solutions for the downlink of a realistic 4[th] generation mobile radio system.

ACKNOWLEDGMENT

The work presented in this paper was partly supported by the European FP5 IST project MATRICE (Multicarrier-CDMA TRansmission Techniques for Integrated Broadband CEllular Systems – www.ist-matrice.org).